\let\includefigures=\iffalse
%
% the following is to use blackboard bold fonts --
\let\useblackboard=\iftrue
%
% activate this if you don't have them.
%\let\useblackboard=\iffalse
%
% You might also need to remove this line.
\newfam\black
\input harvmac.tex
\includefigures
\message{If you do not have epsf.tex (to include figures),}
\message{change the option at the top of the tex file.}
\input epsf
\def\figin{\epsfcheck\figin}\def\figins{\epsfcheck\figins}
\def\epsfcheck{\ifx\epsfbox\UnDeFiNeD
\message{(NO epsf.tex, FIGURES WILL BE IGNORED)}
\gdef\figin##1{\vskip2in}\gdef\figins##1{\hskip.5in}% blank space instead
\else\message{(FIGURES WILL BE INCLUDED)}%
\gdef\figin##1{##1}\gdef\figins##1{##1}\fi}
\def\DefWarn#1{}
\def\figinsert{\goodbreak\midinsert}
\def\ifig#1#2#3{\DefWarn#1\xdef#1{fig.~\the\figno}
\writedef{#1\leftbracket fig.\noexpand~\the\figno}%
\figinsert\figin{\centerline{#3}}\medskip\centerline{\vbox{\baselineskip12pt
\advance\hsize by -1truein\noindent\footnotefont{\bf Fig.~\the\figno:} #2}}
\bigskip\endinsert\global\advance\figno by1}
%%%
\else
\def\ifig#1#2#3{\xdef#1{fig.~\the\figno}
\writedef{#1\leftbracket fig.\noexpand~\the\figno}%
%\figinsert\figin{\centerline{#3}}\medskip\centerline{\vbox{\baselineskip12pt
%\advance\hsize by -1truein\noindent\footnotefont{\bf Fig.~\the\figno:} #2}}
%\bigskip\endinsert
\global\advance\figno by1}
\fi
\useblackboard
\message{If you do not have msbm (blackboard bold) fonts,}
\message{change the option at the top of the tex file.}
\font\blackboard=msbm10 scaled \magstep1
\font\blackboards=msbm7
\font\blackboardss=msbm5
\textfont\black=\blackboard
\scriptfont\black=\blackboards
\scriptscriptfont\black=\blackboardss

\else

\fi
% *************************************
%\draft
%
\def\yboxit#1#2{\vbox{\hrule height #1 \hbox{\vrule width #1
\vbox{#2}\vrule width #1 }\hrule height #1 }}
\def\fillbox#1{\hbox to #1{\vbox to #1{\vfil}\hfil}}
\def\ybox{{\lower 1.3pt \yboxit{0.4pt}{\fillbox{8pt}}\hskip-0.2pt}}

\def\comments#1{}

\def\p{\partial}

\def\half{{1\over 2}}

\def\tr{{\rm tr\ }}
\def\Re{{\rm Re\hskip0.1em}}

\def\CA{{\cal A}}

\def\CT{{\cal T}}
\def\CM{{\cal M}}
\def\CN{{\cal N}}

\def\CL{{\cal L}}

\def\CW{{\cal W}}

\def\ap{\alpha'}

\def\II{\relax{I\kern-.10em I}}

\def\IZ{\relax\ifmmode\mathchoice
{\hbox{\cmss Z\kern-.4em Z}}{\hbox{\cmss Z\kern-.4em Z}}
{\lower.9pt\hbox{\cmsss Z\kern-.4em Z}}
{\lower1.2pt\hbox{\cmsss Z\kern-.4em Z}}\else{\cmss Z\kern-.4em
Z}\fi}
\def\IB{\relax{\rm I\kern-.18em B}}
\def\IC{{\relax\hbox{$\inbar\kern-.3em{\rm C}$}}}
\def\ID{\relax{\rm I\kern-.18em D}}
\def\IE{\relax{\rm I\kern-.18em E}}
\def\IF{\relax{\rm I\kern-.18em F}}
\def\IG{\relax\hbox{$\inbar\kern-.3em{\rm G}$}}
\def\IGa{\relax\hbox{${\rm I}\kern-.18em\Gamma$}}
\def\IH{\relax{\rm I\kern-.18em H}}
\def\II{\relax{\rm I\kern-.18em I}}
\def\IK{\relax{\rm I\kern-.18em K}}
\def\IP{\relax{\rm I\kern-.18em P}}
%\def\IX{\relax{\rm X\kern-.01em X}}
%this doesn't work

%

\def\inbar{\,\vrule height1.5ex width.4pt depth0pt}

\def\p{\partial}

\font\cmss=cmss10 \font\cmsss=cmss10 at 7pt
\def\IR{\relax{\rm I\kern-.18em R}}

\def\BR{\IR}

\def\BR{\IR}
\def\BC{\IC}

\def\lp10{l_P^{10}}
\def\lp11{l_P^{11}}
\def\R11{R_{11}}

\Title{\vbox{\baselineskip12pt\hbox{hep-th/9703056}
\hbox{RU-97-11}}}
{\vbox{
\centerline{D-Branes in Curved Space} }}
\centerline{Michael R. Douglas}
\medskip
\centerline{Department of Physics and Astronomy}
\centerline{Rutgers University }
\centerline{Piscataway, NJ 08855--0849}
\centerline{\tt mrd@physics.rutgers.edu}
\bigskip
\noindent
We obtain actions for $N$ D-branes occupying points in a manifold
with arbitrary K\"ahler metric.  In one complex dimension, the action
is uniquely determined (up to second order in commutators) by the
requirement that it reproduce the masses of stretched strings and by
imposing supersymmetry.  These conditions are very restrictive in
higher dimensions as well.  The results provide a noncommutative
extension of K\"ahler geometry.

\Date{March 1997}
%\draft
%
\lref\dlp{J.~Dai, R.~G.~Leigh and J.~Polchinski, Mod. Phys. Lett. {\bf A4}
(1989) 2073;
J.~Polchinski, Phys.~Rev.~Lett.~75 (1995) 4724-4727;
hep-th/9510017.}
\lref\leigh{R. Leigh, Mod. Phys. Lett. {\bf A4}, 2767.}
\lref\witten{E. Witten, Nucl. Phys. B443 (1995) 85; hep-th/9503124.}
\lref\dgm{M. R. Douglas, B. Greene, and D. R. Morrison, to appear.}
\lref\dg{M. R. Douglas and B. Greene, work in progress.}
\lref\dos{M. R. Douglas, H. Ooguri and S. H. Shenker,
hep-th/9702203.}
\lref\polgeo{J. Polchinski, hep-th/9611050.}
\lref\egs{M. R. Douglas, hep-th/9612126.}
\lref\bfss{T. Banks, W. Fischler, S. H. Shenker and L. Susskind,
hep-th/9610043.}
\lref\DKPS{M. R. Douglas, D. Kabat, P. Pouliot and S. Shenker,
hep-th/9608024.}
\lref\tseytlin{A. Tseytlin, hep-th/9701125.}
\lref\dli{M. R. Douglas and M. Li, hep-th/9412203.}
\lref\connes{A. Connes, {\it Noncommutative Geometry,} Academic Press, 1994.}
\lref\hull{C. Hull, A. Karlhede, U. Lindstr\"om and M. Ro{\v c}ek,
Nucl. Phys. B266 (1986) 1.}
\lref\bl{V. Balasubramanian and F. Larsen, hep-th/9703039.}
\lref\vafa{C. Vafa, hep-th/9602022.}
\lref\taylor{W. Taylor, hep-th/9611042.}
\lref\dewit{B. de Wit, J. Hoppe and H. Nicolai,
Nucl.Phys. B 305 [FS 23] (1988) 545.}
\lref\toappear{M. R. Douglas, A. Kato and H. Ooguri, to appear.}
%
% forward equation references
%
\newsec{Introduction}

In this note we give some results for the world-volume
theories of D-branes in curved space.
D-branes were first defined in weakly coupled
string theory terms \dlp,
and as such their world-volume action is defined by computations
in two-dimensional world-sheet field theory \leigh.
Their importance stretches beyond this; for example they are the
lightest states in the M theory limit \refs{\witten,\bfss}.
Computations in
weakly coupled string theory are not obviously a good way to get at their
physics in other limits, and in situations where the extrapolation
from weak to strong coupling is not determined by supersymmetry
numerous subtleties seem to be emerging, for example in
\refs{\dos,\dg,\bl}.

Here, we approach the problem
as a purely mathematical one: namely, given a particular manifold
with metric $\CM$, we describe a class of $U(N)$ gauge theories
with classical
moduli space $\CM^N/S_N$, such that the small fluctuations have
the expected masses of strings stretched between branes,
in other words proportional to the length of the shortest geodesic
between the branes.
It was noticed in examples in \dos\ that the second condition does not
follow from the first.
As we will see, a solution to this problem defines an interesting
noncommutative analog of Riemannian geometry.
(We might call it ``D-geometry.'')

In general this problem is underconstrained as stated;
computations in string theory (or further consistency conditions)
are required to get a unique answer.  
However, requiring
supersymmetry for the D-brane world-volume action brings additional
constraints.
It turns out that four real supersymmetries ($d=4$, $\CN=1$ supersymmetry),
for which the metric must be K\"ahler,
are enough to get interesting results.

We begin with the simplest non-trivial case -- a curved target space
having two real dimensions, or (by supersymmetry) one complex, and
we show that the constraints have a unique solution, determining the
terms in the action with up to two commutators.

We then discuss higher complex dimensions.
Given a superpotential, we again find strong
constraints on the K\"ahler potential.
This case will be discussed in detail in \toappear.

Finally, as a first step in making contact with relevant mathematics,
we make a coordinate-free definition of the algebra of gauge-invariant
functions on the D-brane configuration space.

\newsec{Kinematics}

We consider a complex manifold $\CM$ with coordinates $z^\mu$ and
K\"ahler potential $K_0(z,\bar z)$.
A single D-brane sitting at a point in $\CM$ and extended in $\BR^k$
is described by the supersymmetrized
Nambu-Born-Infeld action.  In the low energy,
low field strength ($\ap F << 1$ and $|\partial z| << 1$ )
limit this reduces to decoupled supersymmetric sigma model
and $U(1)$ gauge theory Lagrangians
\eqn\nambu{
\CL = \int d^4\theta\ K_0(z,\bar z) + \sum_\mu |X^\alpha|^2
+ \Re \int d^2\theta\ W^2.
}
The flat coordinates $X^\alpha$ play no role in our discussion
and we henceforth drop them.

To describe $N$ D-branes on $\CM$, we promote the coordinates
$z^\mu$ to $N\times N$ matrices $Z^\mu_{ij}$, whose components
are coordinates on a complex manifold $\CM_N$.
We define $\bar Z^\mu = Z^{\mu+}$.
As in flat space, the moduli space modulo gauge transformations
will be parameterized by diagonal matrices whose eigenvalues give the
positions of $N$ branes.
The off-diagonal component $Z^\mu_{ij}$ is a superfield
whose excitations are strings stretched between branes $i$ and $j$.
Each string has a spin (from fermion zero modes) labeled by
the index $\mu$ and component within the superfield.

For manifolds with non-trivial fundamental group $\Gamma$, we would want
to allow stretched strings for every element of $\Gamma$, as in
the toroidal case \refs{\dlp,\taylor,\bfss}.
However, we will only consider a single coordinate patch here.

We will take the $U(N)$ gauge action to be
\eqn\gaugeact{
Z^\mu \rightarrow U^+ Z^\mu U
}
which implies $\bar Z^\mu \rightarrow U^+ \bar Z^\mu U$.
At least locally, this should not be
regarded as an assumption but rather as a choice of coordinate system
for the off-diagonal elements of $Z$.  What we know a priori about the
$U(N)$ action is that it generates orbits generically isomorphic
to the orbits of \gaugeact, and in one complex dimension, we could
instead take as coordinates the D-brane positions and coordinates
on the orbit of the complexified gauge group, e.g. $Z = g^+ z_i \delta_{ij} g$,
for which \gaugeact\ is clear.
In higher dimensions, we are assuming
a result similar to the Frobenius theorem.

The precise definition of the off-diagonal components provided
by \gaugeact\
depends on the choice of coordinate system, and we will return to
this point.

The action will be written in terms of gauge invariant functions,
which are products of traces of products of the matrices $Z^\mu$ and
$\bar Z^\mu$.
The classical action produced by
string theory will satisfy a stronger
condition -- each term will be a single trace of a product of matrices.
This is well-known and follows from the definition of Chan-Paton factors
and the disk topology of the world-sheet.  It is a non-trivial constraint
and was used for example in \tseytlin\ to restrict the possibilities
for the non-abelian Born-Infeld action.
The single trace condition is also required in the application of \bfss, to get
a sensible large $N$ limit describing free particles.

Another evident property of the action derived from the string theory
is that,
considered as a function of formal variables $Z^\mu$ and $\bar Z^\mu$,
the action
takes the same form for any $N$.  Indeed, we do not need to specify $N$
to compute a specific amplitude.
Now the answer for larger $N$ clearly determines the answer for smaller $N$
in numerous ways -- we can take one D-brane far from the others; we can group
the D-branes in pairs and derive an action for $N/2$ subsystems, and so on.
This ``stability'' guarantees consistency under these reductions in $N$.
Thus we can regard the coordinates $Z^\mu$ and $\bar Z^\mu$ as free variables
satisfying no relations, as in \dli.
The action will be largely determined by
a single function of free variables, the K\"ahler potential $\tr K(Z,\bar Z)$.

On the other hand, there is a clear sense in which commutators
$[Z^\mu,\bar Z^\nu]$ are subleading
in the action.  This is essentially to say that we can count
the number of stretched strings in any given process.
The considerations here only involve a single stretched string and
will only determine $K$ up to terms involving two commutators.
It will be very interesting to go further by computing (or postulating)
multistring interactions and proposing an action which summarizes these.

\newsec{D-brane action in one complex dimension}

We begin with this case to illustrate the ideas.
It is not clear to us whether the result will have a physical
interpretation in superstring theory, but we defer discussion of this point
to the end of the section.

The low energy Lagrangian for $N$ D-branes at points in $\CM$ is
\eqn\Dbranes{
\CL = \int d^4\theta\ K(Z,\bar Z)
+ \Re\int d^2\theta\ \tr W^2
}
where $K$ is a single trace $\tr K_N(Z,\bar Z)$.

We can exclude the possibility of a superpotential.
By holomorphy, gauge invariance and the single trace condition,
the only superpotentials we can write are $\CW = \tr \CW(Z)$, and the
corresponding conditions on supersymmetric vacua $\CW'(Z)=0$ would only
have isolated solutions.

A more general gauge kinetic term would have been
$\sum_i \int d^2\theta\ \tr a_i(Z)W b_i(Z)W$.
Assuming that well separated branes are described by \nambu,
the gauge kinetic term for each brane in this limit
is the trivial $\int d^2\theta\ W^2$.
Combining this with holomorphy would then determine the kinetic term
of \Dbranes.
We comment on possible generalizations at the end.

Reproducing the metric for each of the $N$ branes
requires that the K\"ahler potential 
on diagonal matrices $Z_D$ with eigenvalues $z_i$ be
$K(Z_D,\bar Z_D) = \sum_i K_0(z_i,\bar z_i)$.
Thus $K$ must have the same expansion in powers of $Z$ and $\bar Z$
as $K_0$, but with some precise ordering for the $Z$'s and $\bar Z$'s
in each factor.
We will express this by choosing a standard ordering for 
$\tr K_0(Z,\bar Z)$, and represent other possibilities
by adding terms with commutators.

As mentioned above, the terms with up to two commutators will be constrained
by the condition on masses of stretched strings.
The reason they can be determined without doing string
theory computations is that the potential comes entirely
from $D$-terms, and thus these masses
are entirely determined by the gauge action
and the choice of $K$.

Supersymmetry requires the potential to take the form
\eqn\dpot{V = \tr D^2}
with \hull
\eqn\Dterm{
D = [Z,{\p K\over \p Z}] = - [\bar Z,{\p K\over \p \bar Z}].
}
The equality is guaranteed by gauge invariance of $K$.

The masses are then
\eqn\masses{
m_{ij}^2 = {{\p^2 V\over \p Z_{ij}\p \bar Z_{ji}} /
{\p^2 K\over \p Z_{ij}\p \bar Z_{ji}}}
}
for each $i$ and $j$ with no summation implied.  
(One could take $N=2$, $i=1$ and $j=2$ to do this computation).
Using $D=0$ at a supersymmetric minimum, this is
\eqn\massestwo{
m_{ij}^2 = {\tr {\p D\over \p Z_{ij}} {\p D\over \p \bar Z_{ji}} /
{\p^2 K\over \p Z_{ij}\p \bar Z_{ji}}}.
}
Call the denominator of this expression $\delta\bar \delta K$.

We now show that, no matter what ordering we choose for $K$, we have
\eqn\result{
{\p D\over \p Z_{ij}} = (\bar Z_{ii}-\bar Z_{jj})
{\p^2 K\over \p Z_{ij}\p \bar Z_{ji}}.
}
Consider a term in $K$ with a specific ordering.  The derivatives
will produce a sum of terms, one for each appearance of a $Z$ and
each appearance of a $\bar Z$.  Consider for example
\eqn\Kterm{
K = \tr Z \bar Z~ \delta Z~ \bar Z \bar Z Z~ \bar \delta \bar Z~ Z
}
where $\delta$ and $\bar \delta$ mark the appearance on which the partial
derivatives will act.  Since we are working around diagonal $Z$, this
contribution is
\eqn\Ktermval{
{\p^2 K\over \p Z_{ij}\p \bar Z_{ji}} =
 Z_{ii} \bar Z_{ii} ~~ \bar Z_{jj} \bar Z_{jj} Z_{jj} ~~ Z_{ii}.
}
In general, $A \delta Z B \bar \delta \bar Z$
contributes $A(z_i,\bar z_i) B(z_j,\bar z_j)$ to $\delta\bar \delta K$.

Now ${\p D / \p Z_{ij}}$ is also a sum over second derivatives.
Using the representation $D= -\tr [\bar Z,{\p K\over \p \bar Z}]$,
it is easy to see that each term in $K$ produces a corresponding
term in $D$ with the form \result.  The claim follows.

We thus have $m^2_{ij} = |z_i-z_j|^2 \delta\bar \delta K$ and
to reproduce the masses, we require
\eqn\require{
\delta\bar \delta K = { d^2(z_i,z_j) \over |z_i-z_j|^2 },
}
a simple ``correction factor'' to the flat space kinetic
term.  It is non-singular and approaches $\p\bar \p K(z_j)$
as $z_i-z_j\rightarrow 0$.

We now show that there exists an ordering for $K$ which will
reproduce any desired $\delta\bar \delta K$.
Consider an expansion with terms
\eqn\Kexpand{
K = \tr K_0(Z, \bar Z) +
\sum_a \tr f_a(Z,\bar Z) [Z,\bar Z] g_a(Z,\bar Z) [Z,\bar Z] + \ldots .
}
where we take the convention that functions of two variables
are ordered with all $Z$'s before all $\bar Z$'s,
for example
\eqn\Kordering{K_0(Z, \bar Z) = \sum_{a,b} k_{a,b} Z^a \bar Z^b.
}
Terms with more commutators do not contribute to
the second variation.  Consider ${\p^2 K/ \p Z_{12}\p \bar Z_{21}}$,
which is
\eqn\Ksecondvar{\eqalign{
\delta\bar \delta K &=
\sum_{a,b} k_{a,b} {z_1^a - z_2^a\over z_1-z_2}
        {\bar z_1^b - \bar z_2^b\over \bar z_1-\bar z_2} \cr
&\ +
\sum_a |z_1-z_2|^2 \left( f_a(z_1,\bar z_1) g_a(z_2,\bar z_2) +
f_a(z_2,\bar z_2) g_a(z_1,\bar z_1) \right) \cr
&= {1\over |z_1-z_2|^2} \bigg( K_0(z_1,\bar z_1) - K_0(z_1,\bar z_2)
        - K_0(z_2,\bar z_1) + K_0(z_2,\bar z_2) \bigg)\cr
&\ + |z_1-z_2|^2 h(z_1,\bar z_1,z_2,\bar z_2)
}}
where $h$ is a general function invariant under
$z_1\leftrightarrow z_2$.
For small $\epsilon=z_2-z_1$, this has the expansion
\eqn\Ksecondexp{
\delta\bar \delta K = \p\bar \p K(z_1) + O(\epsilon^2)
}
and the $O(\epsilon^2)$ and higher terms are freely adjustable.

We thus have shown that there exists a K\"ahler potential in \Dbranes\ which
satisfies the requirements.  Furthermore, it is uniquely determined up
to terms with more than two commutators.  This is to say that any
term in \Kexpand\ whose second variation is zero,
could be written as a product with more than two commutators.
This can be checked using \Ksecondvar.

It is easy to check that adding dependence on the longitudinal coordinates
$X^\alpha$
in the obvious way (dimensionally reducing a higher dimensional
world-volume theory) works as it should, because $\delta\bar\delta K$
also multiplies the potential in this case.

We now consider the case of non-trivial gauge kinetic term,
as could be produced by a non-constant dilaton background.
The D-term potential becomes
\eqn\dpotf{
V = D (\delta\bar\delta f)^{-1} D
}
where $\delta\bar\delta f$ is the second variation
of the gauge kinetic term with respect to $W$.
This is a diagonal matrix and easy to invert, so this leads to
\eqn\massesf{
m^2_{ij} = |z_i-z_j|^2
{\delta\bar \delta K(z_i,z_j) \over \delta\bar\delta f(z_i,z_j)}.
}
One can check that the gauge boson masses are given by the same
formula.

\subsec{Example -- the two-sphere}

A K\"ahler potential producing the rotationally symmetric metric
in the usual stereographic coordinates is
\eqn\kahlerstwo{
K = \log (1 + z\bar z).
}
The shortest geodesic distance between two points is
\eqn\diststwo{
d(z_1,z_2) = 2 \arctan
{|z_1-z_2|\over (1+z_1 \bar z_2)^{1/2}(1+z_2 \bar z_1)^{1/2}}.
}
Using \require\ and \Ksecondvar\ we have
\eqn\corrstwo{\eqalign{
h(z_1,z_2) &=  { 1 \over |z_1-z_2|^4 }
\bigg( d^2(z_1,z_2) -K_0(z_1,\bar z_1) + K_0(z_1,\bar z_2)
        + K_0(z_2,\bar z_1) - K_0(z_2,\bar z_2) \bigg)\cr
&=  { 1 \over |z_1-z_2|^4 }
\left( d^2(z_1,z_2) - \log {(1+|z_1|^2)(1+|z_2|^2)\over
                                (1+z_1 \bar z_2)(1+z_2 \bar z_1)}\right) \cr
&= { 1 \over |z_1-z_2|^4 }
\left( \log^2 {1+i\sqrt{u}\over 1-i\sqrt{u}} - \log (1+u) \right).
}}
The parenthesized expression has a Taylor expansion in
$u=|z_1-z_2|^2/(1+z_1 \bar z_2)(1+z_2 \bar z_1)$,
so this can be reproduced by \Kexpand.

Explicitly reproducing just the terms up to two commutators in this
way leads to a complicated and unenlightening expression.  It seems
likely that choices exist for the higher commutator terms which
lead to a simpler expression, which would help in finding a more
geometric description of the result.
However, the main point
we want to make at present is that the result is determined.

The physics of the result is in the mixed components of
the curvature on the moduli
space.  These are the second derivatives
of the logarithm of the geodesic distance
$\partial^2 \log d^2(z_i,z_j) / \partial z_i \partial \bar z_j$
(we derive this below).  They could be seen in the low energy
scattering of transverse ripples on the brane.

On general grounds one expects this curvature
to be non-singular except at special points in configuration space
where massless degrees of freedom appear.  In the present example,
the special points in the moduli space are the points where a pair
of D-branes sit at antipodal points of the sphere, and there is no
longer a unique shortest geodesic connecting them.  The resulting
singularity in the curvature should be associated with the
stretched string having acquiring a zero mode for rotations around
the sphere.

\subsec{Physical interpretation}

Although the mathematical
problem is well-defined in one complex dimension, 
and serves to illustrate the general case,
it is not clear whether the result can be
directly interpreted as coming from a string theory.

First of all, these spaces (except for flat space)
are not Ricci flat and these are not solutions of superstring
theory or of M theory.  
Conceivably, it might be
possible to define a classical D-brane action in such background,
by requiring conformal invariance only for the boundary
interactions. 

A more serious problem is that there will be no covariantly constant spinor
on $\CM$, so string theory D-branes on $\CM\times \BR^k$
will not have $\CN=1$ supersymmetry.  Rather than derive fermions
by dimensional reduction, we have implicitly postulated fermions
which are supersymmetry partners of the bosons.

The result for the bosonic part of the action
does seem physically sensible,
and to get this the supersymmetry is being used only as a device
(indeed any two-dimensional Riemannian manifold admits complex
coordinates in which the metric is K\"ahler), so it is quite
possible that the result itself does not depend on supersymmetry, only this
derivation.

Another one-dimensional problem (currently under investigation)
which should have a superstring interpretation
would be to allow a non-constant dilaton background, as in F theory \vafa.

\newsec{General covariance}

The K\"ahler potential \require\ for the off-diagonal modes
is not manifestly covariant -- it depends on the choice of
coordinate $z$.  This can be traced back
to the definition of the gauge action \gaugeact.

On the other hand, physical quantities are covariant.  The masses of
states are, by assumption.  Let us check the Riemann curvature on the
moduli space.  The mixed components at $\phi=0$
(let $\phi$ be an off-diagonal component) are
\eqn\curve{\eqalign{
R_{\bar z z \phi}^{~~~\phi} &=
\p_{\bar z} (g^{\phi\bar\phi} \p_z g_{\phi\bar\phi})
\cr
&= \p_{\bar z} \p_z \log \delta\bar\delta K \cr
&= \p_{\bar z} \p_z \left( \log d^2(z_1,z_2) - \log |z_1-z_2|^2 \right) \cr
&= \p_{\bar z} \p_z \log d^2(z_1,z_2)
}}
except at $z_1=z_2$, where the $\log |z_1-z_2|^2$ serves to cancel
the short distance singularity.
The components $R_{\phi\phi}^{\phi\phi}$, or the curvature
at $\phi\ne 0$, require knowing terms with more commutators to compute.
Thus, to the extent we have computed the action here, it is covariant.

One way to restore manifest covariance,
at generic points in moduli space, would be to absorb the
coordinate dependence into the fields, by making the
field redefinition $\tilde Z_{ij} =  Z_{ij} / (Z_{ii} - Z_{jj})$.
However this breaks down when $Z_{ii} = Z_{jj}$ and is obviously not
a good definition.

A better way to implement covariance is to
postulate non-trivial transformation laws for the
off-diagonal components of $Z$.  Evidently they should transform like
differences of coordinates.

The simplest treatment would be to postulate a transformation law
for the entire matrix $Z$.  In one dimension, there is only one
possibility for a holomorphic coordinate transformation
compatible with gauge invariance.  We can only write
\eqn\noncd{ Z = f(Z') }
where $z = f(z')$ is the coordinate transformation on $z$.

Thus we ask whether the expression \Kexpand\ is covariant under
this definition of change of coordinate.  A nice feature of the
ordering prescription we used is that it is preserved under \noncd,
so the functions transform as
\eqn\transform{
K_0(Z,\bar Z) \rightarrow K_0(f(Z'),f^*(\bar Z'))
}
and so on.
The second variation \Ksecondvar\ with respect to the off-diagonal
components of $Z'$ becomes
\eqn\Ksecondtwo{\eqalign{
{\p^2 K'\over \p Z'_{12}\p \bar Z'_{21}}
&= {1\over |z'_1-z'_2|^2} \bigg( K_0(f(z'_1),f^*(\bar z'_1))
        - K_0(f(z'_1),f^*(\bar z'_2)) \cr
&\qquad\qquad\qquad     - K_0(f(z'_2),f^*(\bar z'_1))
        + K_0(f(z'_2),f^*(\bar z'_2)) \bigg)\cr
&\ + {|f(z'_1)-f(z'_2)|^4 \over |z'_1-z'_2|^2}
 h(f(z'_1),f^*(\bar z'_1),f(z'_2),f^*(\bar z'_2)) \cr
&= {|z_1 - z_2|^2 \over |z'_1-z'_2|^2}
{\p^2 K\over \p Z_{12}\p \bar Z_{21}}
}}
and the masses $m^2_{12} = |z'_1-z'_2|^2 \delta'\bar\delta' K'$
are invariant.

Since the action was determined by the fact that it reproduced the
metric and masses, and these transform properly, we conclude that
this definition of the action is indeed covariant under the simple extension
of change of coordinates \noncd.

\newsec{Higher dimensions}

We introduce complex matrix
coordinates $Z^\mu$ with $1\le \mu\le D$ and
make no restriction on $D$.
The strategy will again be to reproduce the masses of stretched
strings by finding some correct ordering of the K\"ahler potential
$K(Z,\bar Z)$.
We take the masses for
every polarization of the stretched string to be $m^2=d^2$.
This seems quite plausible at least at leading order in $\ap$
in string theory, since these differ only
in the state of their fermion zero modes.

The D terms are $D=\sum_\mu [Z^\mu,\p_\mu K]$ and $D=0$ combined with
gauge quotient will leave a moduli space of complex dimension $(d-1)N^2+N$
Thus a superpotential is required to restrict the moduli space
to commuting $[Z^\mu,Z^\nu]=0$.  This can be accomplished by
a generic superpotential of the form
\eqn\superp{
W = \half\sum_{\mu,\nu} \tr  w_{\mu\nu}(Z) [Z^\mu, Z^\nu].
}

Computing the mass matrix for the off-diagonal fields $Z_{12}^\mu$
will again involve their variations,
$\delta_\mu X = {\p X/ \p Z^{\mu}_{12}}$.
One finds
\eqn\masshigh{
m^2_{1\mu,2\nu} = (\delta\bar\delta K)^{-1,\mu\bar\rho} \left(
\bar\delta_{\bar\rho} D \delta_\nu D +
(\delta_\lambda\bar\delta_{\bar\lambda'} K)^{-1}
\bar\delta_{\bar\rho}\bar\p_{\bar\lambda'} \bar W~
        \delta_{\nu}\p_\lambda W \right).
}

The result \result\ generalizes in an obvious way:
\eqn\highresult{
\delta_\mu D = \sum_{\bar\nu}
(\bar Z^{\bar\nu}_{11}-\bar Z^{\bar\nu}_{22})
\delta_\mu \bar\delta_{\bar\nu} K.
}

Explicit formulas for $\delta_\mu \bar\delta_{\bar\nu} K$
and $\delta_\nu\p_\lambda W$ are complicated by the fact that
now we have to choose an ordering among the holomorphic coordinates.
One possibility is to totally symmetrize, writing
\eqn\Kstuff{
K =
\oint \prod_\mu {d\alpha^\mu\over \alpha^\mu}~
\oint \prod_\mu {d\bar\alpha^\mu\over \bar\alpha^\mu}~
\tr {1\over 1-\sum_\mu Z^\mu/\alpha^\mu}
{1\over 1-\sum_\mu \bar Z^\mu/\bar\alpha^\mu}~
K_0(\alpha,\bar\alpha)
}
as the leading term, and then expressing corrections to this in
terms of commutators.  However this is not preserved by
holomorphic coordinate transformations and it is not obvious
that it is natural, or indeed that there is any natural
universal ordering.  Thus we refrain from writing explicit analogs
of \Kexpand\ and \Ksecondvar, and content ourselves with the
observation that there is again enough freedom to produce
\eqn\Karb{
\delta_\mu\bar\delta_{\bar\nu} K = \p_\mu\bar\p_{\bar\nu} K_0 +
(z_1^\rho-z_2^\rho)(\bar z_1^\lambda-\bar z_2^\lambda)
h_{\mu\rho\nu\lambda}(z_1,z_2)
}
with $h_{\mu\nu\rho\lambda}(z_1,z_2)$ symmetric in $z_1\leftrightarrow z_2$
and in $\mu\nu\leftrightarrow\rho\lambda$ but otherwise arbitrary,
and
\eqn\Warb{
\delta_\nu\p_\lambda W = (z_1^\rho-z_2^\rho) w_{\nu\lambda\rho}(z_1,z_2)
}
with $w$ a holomorphic function satisfying
$w_{\nu\lambda\rho}(z_1,z_2)= -w_{\lambda\nu\rho}(z_2,z_1)$
and $(z_1^\nu-z_2^\nu)w_{\nu\lambda\rho}(z_1,z_2)=0$
(this follows from gauge invariance).

Thus, $\delta\bar\delta K$ already has enough freedom to reproduce
the masses.  The freedom in $W$ is not enough to do it alone but
does make the answer non-unique, in a way analogous to the non-uniqueness
we found in one dimension if we introduced a general gauge kinetic term.

Of course we might have other physical constraints on the superpotential,
for example that it be non-singular, which would be very constraining
if $\CM$ is compact.  Another interesting case is $\CN=2$, $d=4$
supersymmetry, which requires $\CM$ to be hyperk\"ahler, and
for which the superpotential is also uniquely determined \hull.
This suggests the possibility that the action is uniquely determined
up to higher commutators in this case as well.

Additional knowledge about the K\"ahler potential would also fix this.
For example, we might try the ansatz
``$\delta_\mu\bar\delta_{\nu} K \propto g_{\mu\bar\nu}$''.
This is not sensible as it stands -- we need a quantity depending both
on $z_i$ and $z_j$, which is likely to be an integral of a locally
defined tensor along the geodesic.
This is a point at which computation in superstring theory
(or other underlying definition of D-brane) might be required to completely
determine the action.  
One limit to which this action naturally applies is weak curvature
$\ap R << 1$ and long stretched strings, $|z_i-z_j|^2 >> \ap$, and
computations in this regime do not look difficult, but
we leave this for future work.

\newsec{Noncommutative geometry}

Clearly we are talking about some sort of ``noncommutative geometry,''
and it will be interesting to make contact with related work
in mathematics, such as that of Connes \connes.

As a start, let us define the algebra of gauge-invariant functions
which we used here in a coordinate-free way.
It would be very useful to have
a coordinate-free version of the present discussion.

Let $M_N(\BC)$ be the algebra of complex
$N\times N$ matrices, and $\CM_N$ the $DN^2$-dimensional configuration
space of the D-brane theory admitting a $U(N)$ action $\pi$ with
the general properties above.
Then $\CA_N$ is the subalgebra of
$ M_N(\BC) \otimes C^\infty(\CM_N)$ satisfying
\eqn\equi{
g^{-1} a g = \pi^*(g) a.
}
In our earlier discussion, the action $\pi^*$ was given by \gaugeact,
and we used the statement that $\CA_N$ was generated by $Z$ and $\bar Z$.
The action is a $U(N)$ invariant linear functional of an element of
$\CA_N$, i.e. a trace.

Another application of a precise definition would be to facilitate
taking the large $N$ limit -- we would just replace $M_N(\BC)$
with another algebra $\CT$.  We also need to replace $\CM_N$ with
a space which locally is modelled on $\CT \otimes \BR^D$.
This would be useful (for example) in describing membranes along the
lines of \refs{\dewit,\bfss}.

\smallskip
I would like to thank Costas Bachas, Brian Greene, Akishi Kato,
Hirosi Ooguri, John Schwarz and Steve Shenker for discussions,
and the Caltech theoretical physics group for their
hospitality.
This research was supported by DOE grant DE-FG02-96ER40959.

\bigskip

\listrefs
\end